**HIV-1 virus cycle replication: a review of RNA polymerase II transcription, alternative splicing and protein synthesis**

Corresponding autor: Miguel Ramos-Pascual

Abstract

HIV virus replication is a time-related process that includes attachment to host cell and fusion, reverse transcription, integration on host cell DNA, transcription and splicing, multiple mRNA transport, protein synthesis, budding and maturation. Focusing on the core steps, RNA polymerase II transcripts in an early stage pre-mRNA containing regulator proteins (i.e *nef*,*tat*,*rev*,*vif*,*vpr,vpu*), which are completely spliced by the spliceosome complex (0.9kb and 1.8kb) and exported to the ribosome for protein synthesis. These splicing and export processes are regulated by *tat* protein, which binds on Trans-activation response (TAR) element, and by *rev* protein, which binds to the Rev-responsive Element (RRE). As long as these regulators are synthesized, splicing is progressively inhibited (from 4.0kb to 9.0kb) and mRNAs are translated into structural and enzymatic proteins (*env*, *gag-pol*). During this RNAPII scanning and splicing, around 40 different multi-cystronic mRNA have been produced.

Long-read sequencing has been applied to the HIV-1 virus genome (type HXB2CG) with the HIV.pro software, a fortran 90 code for simulating the virus replication cycle, specially RNAPII transcription, exon/intron splicing and ribosome protein synthesis, including the frameshift at *gag*/*pol* gene and the ribosome pause at *env* gene. All HIV-1 virus proteins have been identified as far as other ORFs. As observed, t*at*/*rev* protein regulators have different length depending on the splicing cleavage site: *tat* protein varies from 224aa to a final state of 72aa, whereas *rev* protein from 25aa to 27aa, with a maximum of 119aa. Furthermore, several ORFs coding for small polypeptides sPEP (less than 10 amino acids) and for other unidentified proteins have been localised with unknown functionality.

This review includes other points in the virus cycle that remain still unclear, as future research lines, such as the attachment to other host cells different to CD4 lymphocytes, the reverse-transcription and integration of the second RNA strand of the HIV virus, the presence of several cleavage sites for polyadenylation in the whole genome, regulation of splicing process, other ribosomal pausing/frameshifting, the antisense protein synthesis, docking of gp120/41 glycoproteins into cell-membrane and characterisation of binding site region to CD4 lymphocytes co-receptors CCR5/CXCR4, the analysis of the virus cell membrane, including the presence of CD4 co-receptors or the protease effect during maturation process.

The detailed analysis of the HIV virus replication and the characterisation of virus proteomics are important for identifying which antigens are presented by macrophages to CD4 cells, for localizing reactive epitopes or for creating transfer vectors to develop new HIV vaccines and effective therapies.

## 1. Introduction

HIV (Human Immunodeficiency virus) is a virion of the family Retroviridiae Lentivirus, which stays in a latent state in the host body during several years, developing later, without an effective treatment, the acquired immunodeficiency syndrome (AIDS) [1, 2].

The HIV virus replicates through a special cell of the immunity system, the T-helper lymphocytes cells or CD4. After infection and replication, the total number of these cells reduce considerably, increasing the apparition of other opportunistic diseases, what is known as the AIDS syndrome [3]. As T-helper CD4 cells are the major focus of HIV infection, other cells of the immune system are also attacked by the HIV virus, such as dendritic cells, which are one of the first cell to encounter HIV in the mucosal epithelia [4].

The HIV virus cycle has been extensively studied and analysed, still focus of current research [5], as far as the HIV genomics and proteomics [6]. HIV virus replication is a time-related process that includes host cell fusion, reverse transcription, integration on DNA host cell, transcription, pre-mRNA splicing, multiple mRNA transport, protein synthesis, budding and maturation.

Several studies have focused on the transcription process, including the spliceosome complex assembly and splicing [7-13]. Protein synthesis and protease enzymatic activity is still under research [14]. Others have investigated the geometrical evolution of the capsid during maturation [15,16]. Despite all these advances, some features of the virus replication are still unknown, such as the role of some genes during translation or the presence of further open reading frames (ORF) encoding polypeptide chains, still not identified [17].

RNA structure prediction software has been previously developed and classified based on the method applied to the genome sequence. These codes analyse the genome nucleotide per nucleotide, comparing codons with consensus sequences or repetition sequences [18] The HIV.pro software has been developed for simulating the HIV-1 virus cycle replication, focusing in the transcription, splicing and protein synthesis.

This review includes other points in the virus cycle replication that remain still unclear, such as the attachment to other host cells different to CD4 lymphocytes, the reverse-transcription and integration of the second RNA strand of the HIV virus, the presence of several cleavage sites for polyadenylation in the whole genome, regulation of splicing process, other ribosomal pausing/frameshifting, the antisense and other protein synthesis, docking of gp120/41 glycoproteins into cell-membrane and characterisation of binding site region to CD4 lymphocytes co-receptors CCR5/CXCR4, the analysis of the virus cell membrane, including the presence of CD4 co-receptors or the protease effect during maturation process.

2. The HIV retrovirus: structure, genome and virus cycle

2.1 Structure of the virion

The HIV virus has a diploid single-stranded RNA genome protected by several envelopes: a spherical lipid membrane, a matrix (MA) of protein p17, which support the membrane, a cone-based capsid (CA) of protein p24, and a nucleocapsid (NC) of protein p7, binding to the viral RNA. Spacer peptides p2/p1 are protein connectors making the nucleocapsid stable. Another protein, p6, is auxiliary in the processes of incorporation of the genome into new viruses [14-16].

On the external surface of the spherical capsid, it is localised the glycoprotein gp160, which is the combination of a protuberance spike (gp120), with several conformation loops (V1 to V5), and a transmembrane connector (gp41) [19-20]. Virus membrane is produced through budding of host-cell membrane, so can include host-cell connectors in the surface (i.e. MHCII, CCR5…) [54-55]

Inside all these structures, several enzymes are present, such as reverse-transcriptase (RT), protease (PR) and integrase (IN). All retroviruses undergo reverse-transcription of the messenger mRNA into double-stranded DNA, which once introduced into the host DNA cell, is transcribed into mRNA segments that produce the new proteins of the virus.

Other viruses, such as the poliovirus that causes paralytic poliomyelitis, have similar structures, enzymes and envelopes as the HIV virus [21]

2.2 Virus genome

The HIV-1 genome is composed of a duplicated positive-sense single-stranded RNA. Each RNA strand has approximately 9700 nucleotides long and encodes all the proteins and enzymes of the virus. The genome is poly-adenylated at its 3'-terminus and it contains a long 5'-terminus followed by a short repetition 5'-R terminus, which includes the poly(A) tail. Another special characteristic of this genome is a secondary 5'-terminus after the repetition ltr sequence (see Annex A).

The genome has several genes classified as: structural (*gag/pol/env*), regulatory (*tat/rev*) and accessory (*nef/vif/vpr* and *vpu*). These genes decode envelope virion proteins, create enzymes, and promote the transcription and infection of the virus. Fig. 1 shows the position of the HIV-1 (HXB2) genes in the genome.

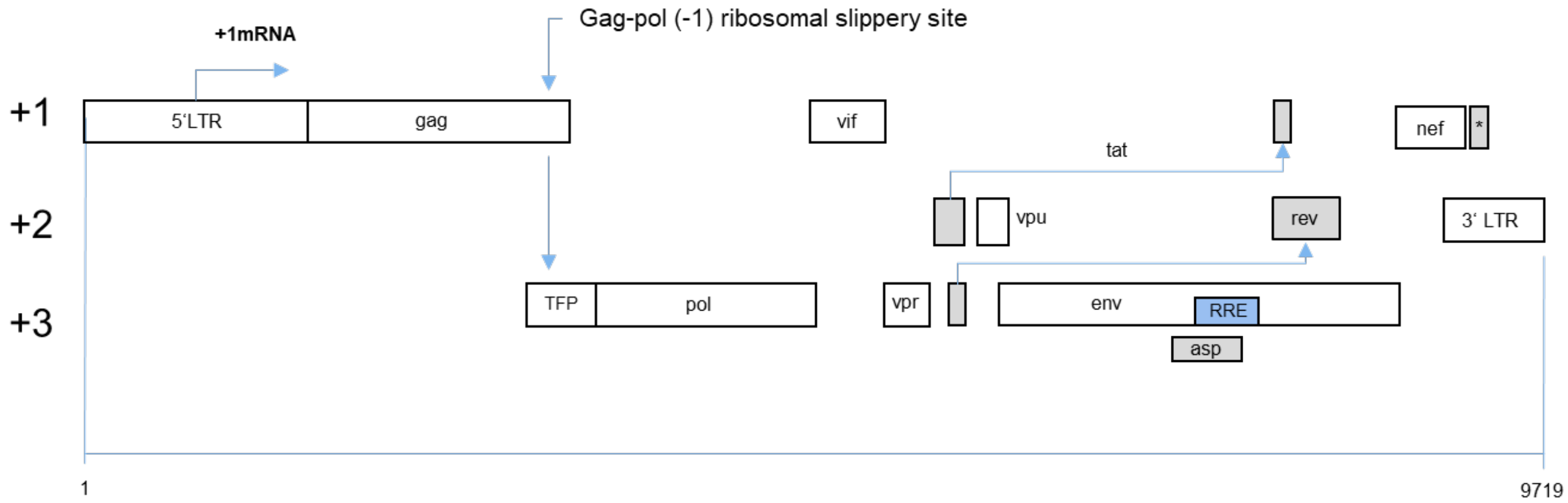

Figure 1 – Scheme of the HIV-1 (HXB2) genes and position in the frameshift

## 2.3 HIV proteins coding sequences (CDS)

HIV genes superpose each other in three different reading frames [22]. An open reading frame (ORF) is defined as any nucleotide sequence starting with an AUG- codon and ending with codon -UAA/UAG/UGA, which can be translated into a protein in the ribosome. If a sequence contains several ORF, the mRNA fragment is called multi-cystronic.

HIV virus is polycistronic, that is, messenger RNAs encode multiple polypeptides in a segment. For example, the *gag-pol* or the *vpu-env* genes produce bicistronic mRNA [23-27] The second (and subsequent) ORF are less efficiently translated by the ribosome, unless an IRES (Internal Ribosome Entry Site) is found between them [28-30] [62]. Table 1 presents the main genes and a description of the function during the replication.

Table 1 - Summary of the main genes and characteristics of HIV-1 [31] [47]

| Type | Gene | Function |
|---|---|---|
| Structural | gag | Envelope of the virion: MA (p17), CA (p24), NC (p7), p2/p1, p6 |
| | Pol | Enzymes protease (p10), reverse-transcriptase (p51/p66), R nuclease (p15) and integrase (p31) |
| | Env | Protein gp160 on the virion envelope: Spike glycoprotein (gp120) and transmembrane region (gp41) |
| Regulatory | Tat | Promotor of the transcription of viral genome – enhances RNA polymerase II mediated elongation |
| | Rev | Promotor of export of incompletely spliced viral RNAs |
| | Vif | Virion infectivity factor p23 – inhibitory effects of host factor (APOBEC3G) |
| | Vpr | Viral protein R – increases viral replication, facilitates infection of macrophages |
| | Vpu | Viral protein U – promotes CD4 degradation and virion release |
| | Nef | Negative effector p24 – promotes downregulation of surface CD4 and MHCI expression, promotes viral infectivity |
| Others | Asp | Antisense Protein (ASP), citation in [65,66] |

## 2.4 Virus replication cycle

### 2.4.1 Introduction

HIV virus replication is a time-related process that includes attachment to host cell and fusion, reverse transcription, integration on host cell DNA, transcription and splicing, multiple mRNA transport, protein synthesis, budding and maturation (see figure 2)

### 2.4.2 Attachment and fusion

HIV virus binds preferentially to CD4 lymphocytes, specially to the CD4 receptor in the cell membrane, creating a change in the conformation of the gp120 glycoprotein loops and binding to the chemokine co-receptor CCR5 or co-receptor CXCR4. After this attachment, transmembrane gp41 produces fusion of both membranes, releasing provirus into the cytosol [32].

As HIV virus membrane is formed from CD4 lymphocytes cell membrane, other co-receptors (i.e. CCR5 and CXCR4) could be present and used by the virus during the infection process [64].

Some other studies have proved binding with other cells of the immune system, as dendritic cells, as commented previously [4].

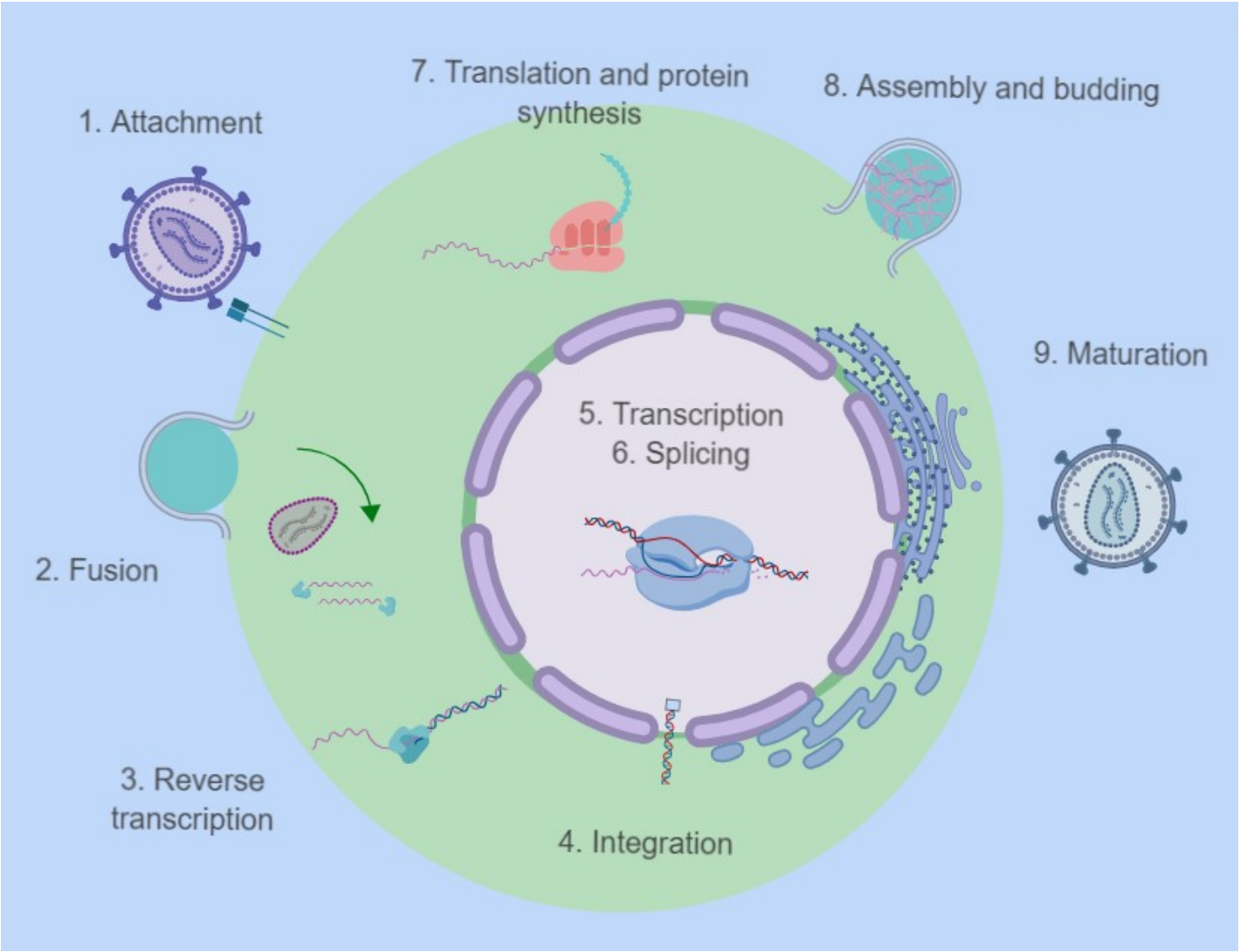

Figure 2 – Scheme of the HIV virus cycle replication (Image generated with BioRender http://app.biorender.io)

### 2.4.3 Reverse-transcription and integration

The single-stranded RNA undergoes reverse-transcription by the virus reverse-transcriptase enzyme, producing first a cDNA strand with virus RNA as template. Ribonuclease H degrades the union between the template RNA and cDNA [33] and later the cDNA is complemented into a dsDNA by the reverse-transcriptase [34]. HIV Integrase enzyme binds the copied virus dsDNA into the host cell DNA, specially into active transcription units, as the LEDGF/p75 factor [35].

If the second +ss RNA of the virus is also reverse-transcribed and integrated has not been previously reported. Figure 3 presents a model of the HIV reverse-transcription with double +ss RNA and double integration, in 1-site with directions 5’-3’, 5’-5’ or 3’3’, or in 2 different sites of the host cell.

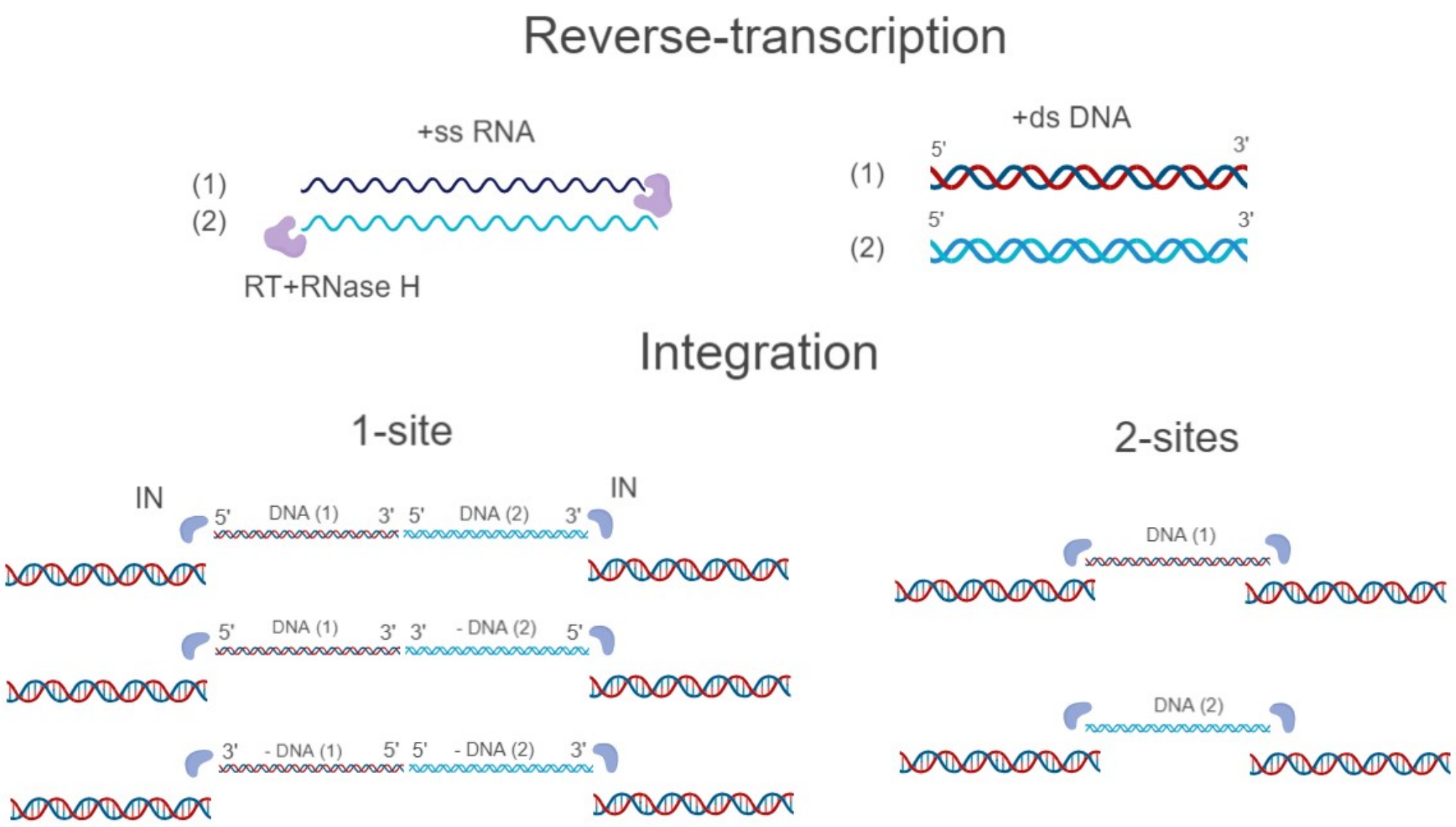

Figure 3 – Scheme of a double reverse-transcription and integration during the HIV virus replication cycle, depending on different hypotheses: 1-site integration, with three different viral DNA directions and 2-sites integration (Image generated with BioRender http://app.biorender.io)

2.4.4 Transcription

RNA polymerase II of the host cell starts to transcript precursor mRNA. RNAPII interacts with specific promoters in the viral DNA, located in the 5'ltr region [36]. The minimal functionality for transcription needs the presence of three elements (promoter region): SP1 binding sites, the TATA box and an initiator sequence. Another promoter region has been identified in the 3'ltr [37-39]

Firstly, and during transcription, a 5'cap is added to the pre-mRNA. After capping, there is an elongation phase, in which sometimes splicing also occur. After that the segment is released and in a competitive way, the 3' terminus is polyadenylated and the sequence spliced.

2.4.5 Splicing

Splicing is the segmentation of the transcribed sequence to a final mRNA, allowing the export of small fragments of mRNA out from the nucleus and producing protein isoforms from the same gene. Exons and introns are delimited by 5' splice donors and 3' acceptors splicing points [7-13] [63]

At the beginning of transcription, precursor mRNA is fully spliced into segments of 1.8kb which pass through the nucleus membrane and are exported to the ribosome. As long as the viral regulatory proteins *rev* and *tat* increase [40-41], the complex process of gene expression is accelerated and full-length mRNA is transported from the nucleus to the cytoplasm.

The length and complexity of the mRNA increases from fully spliced to incompletely spliced, as the splicing process is inhibited, decreasing the efficiency of spliceosomes. *Tat* protein binds with Transactivation Responsive region (TAR), increasing the transcription of pre-mRNA. *Rev* protein interacts with a region on env gene, the Rev Responsive Element (RRE), exporting out from the nucleus the mRNA before being completely spliced. The RRE element is a ~350 nucleotide region in the vpu/env gene which acts as a scaffold of the rev protein. The length of the mRNA segments increases from from 1.8kb to 4.0kb and 9.0kb mRNA lengths. In this late phase, the capsid and envelope proteins are translated and rearrange into an immature virion state [42]. Figure 4 shows a scheme of the complexity of the splicing process depending of the stage.

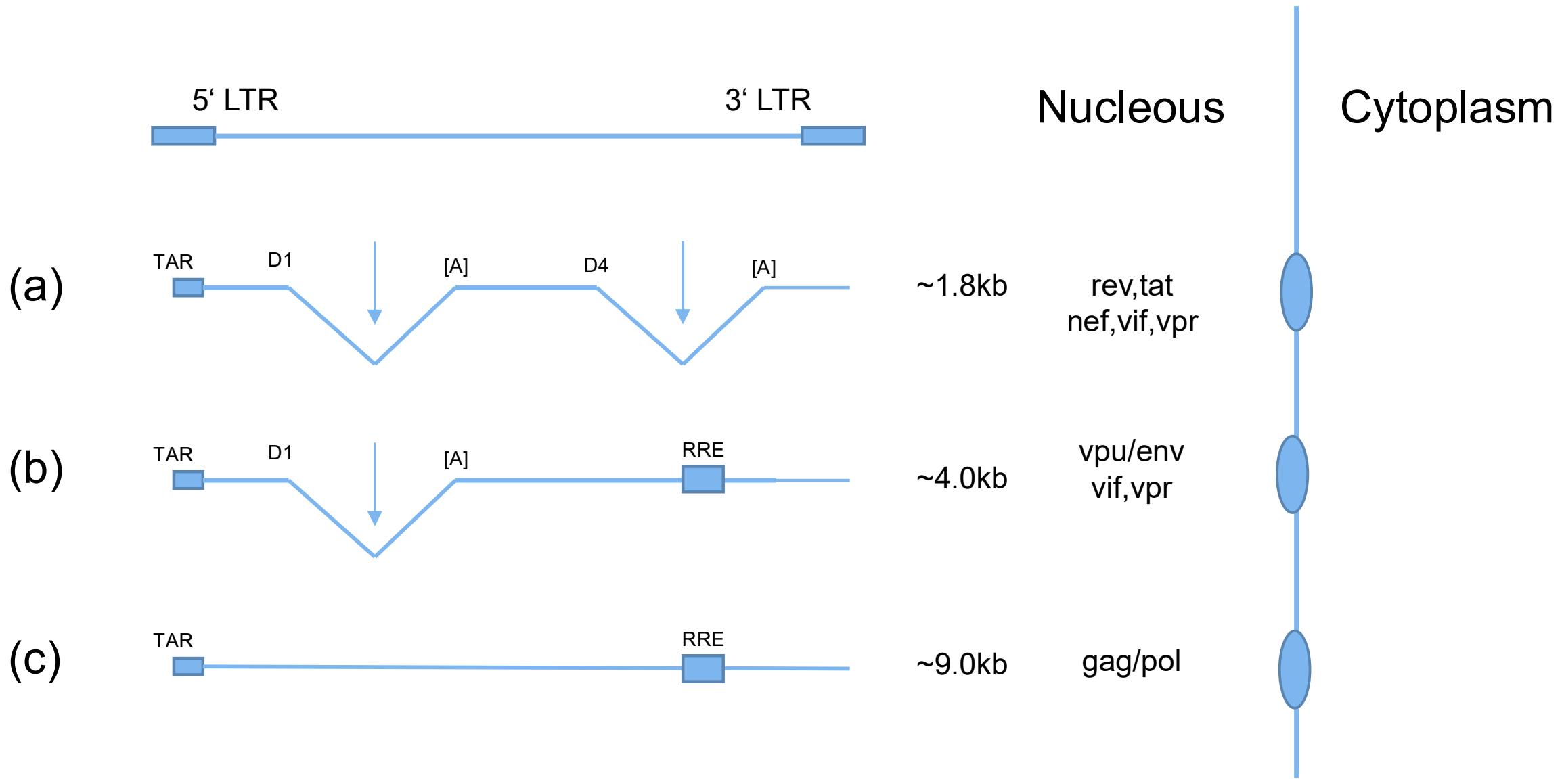

Figure 4 – (a) At an early stage, provirus genome undergoes several inefficient splicing, releasing introns D1[A] and D4[A] which have the highest spliceosome binding intrinsic strengths [56], with [A] a variable 3'ss and producing 1.8kb mRNA, which encode proteins *rev/tat* and *nef,vif,vpr*. (b) In an intermediate stage, as *rev* protein binds with RRE element, 4.0kb mRNA are exported before splicing D4[A] and vpu/env protein is translated. (c) At the end, splicing D1[A] is inhibited and the full-length genome 9kb is exported out from the nucleus and gag/pol protein translated.

2.4.6 Translation, protein synthesis and maturation

Some of the methods that the HIV uses for protein synthesis are frameshifting, ribosome shunting/leaky scanning and cap-independent mechanisms though IRES (Internal Ribosome Binding Site) [23-24] [28-30]. Other methods are SP-(Stem Pause), producing breakage of the chain during peptide synthesis, as the SP-stem located between the spacer peptide 1 and gp120 protein in env gene [6].

HIV genome includes a programmed (-1) ribosomal frameshift, also known as 'slippery site', in the gag-pol gene. During ribosome scanning and after this frameshift, reading frame changes to another and start translation of *pol* protein [43-44] This change in the reading frame occurs around 5-10% of the times, permitting the development of capsid proteins in a frequency higher than virus enzymes.

A low frequency of cases, around 5%, the AUG- codon is skipped and translation starts on the second, which is significative in multi-cystronic mRNAs. The second (and subsequent) ORF are less efficiently translated by the ribosome, unless a IRES (Internal Ribosome Entry Site) is found between them.

Protease enzyme synthetizes these proteins after proteolysis of the long chain polypeptides released by the ribosome [45] Although some studies have reported different cleavages sites in a polypeptide chain for a single protease [46], HIV protease has a high sequence specificity for single cleavage points in octapeptide regions, for instance in the gag-pol fusion protein.

After proteolysis, virion is assembled in an immature state, crosses cell membrane (budding) and further structural changes into the proteins produce the final virus state (maturation) [57].

3. HIV.pro software

3.1 Description

The software HIV.pro is a Fortran f90 code developed for the analysis and simulation of HIV-1 virus and protein synthesis. The software reads as input the virus genome (RNA) from an external file in FASTA format, and using a single long-term sequencing method, after reverse-transcription and RNAPII transcription, alternative splicing and intron removal, identifies the polypeptide chains in all possible mRNA. Each mRNA has been spliced using the donor and acceptor splicers defined in [10] [47]. Figure 5 presents the main screen out of the software.

The software searches for open reading frames (ORF) in the RNA sequence after cleavage and splicing, considering the reading frame, that is, separated by a three-multiple number of codons and creates an output file for each one of these proteins in FASTA format [48]. The NCBI REFSEQ (Ag 2018) HIV virus sequence has been used as input for the HIV.pro software, as commented previously [31] [58-60].

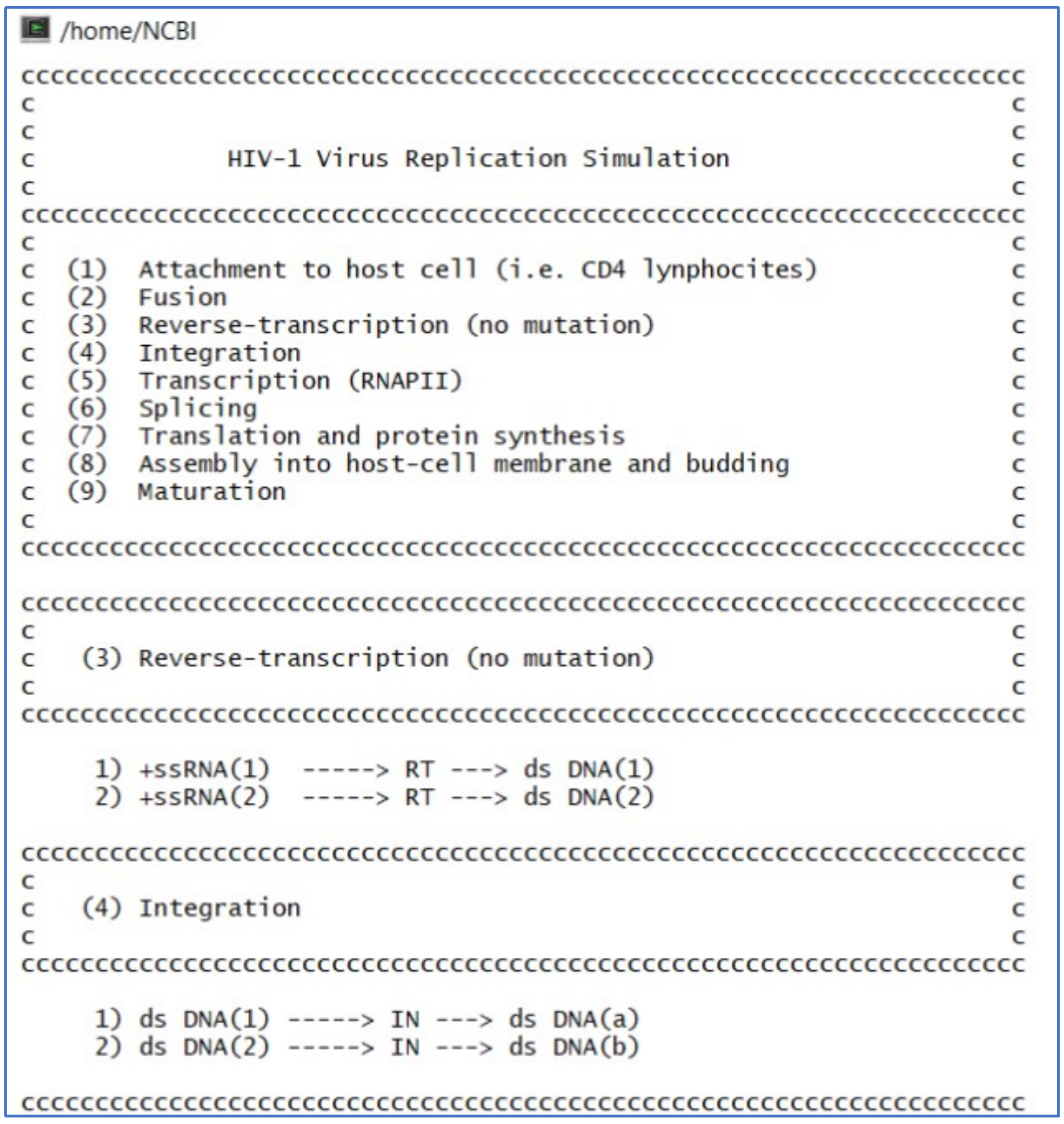

Figure 5 – HIV.pro software for simulating HIV virus replication cycle

### 3.2 Transcription and splicing model

During all splicing process, more than 40 different classes of mRNA are produced with different lengths, from approximately 1.8kb (completely spliced) to 9kb (incompletely spliced). These segments have been modelled with the 5'ss donor and 3'ss acceptor sites from [10]. Tables 2-3 show the details of these mRNA segments and the definition of exon/intron. A4* and A7* have several different cleavage sites for the spliceosome complex protein U2 snRNP, suggesting unstable cleavage points in some stages of the splicing.

Table 2 – Description of the introns 5'and 3' donor/acceptor sites [10]

| 5' splice donor (Γ) | | 3' splice acceptor (Ј) | |
|---|---|---|---|
| D0 | 1 | A0 | 455 |
| D1 | 743 | A1 | 4913 |
| D2 | 4962 | A2 | 5390 |
| D3 | 5463 | A3 | 5777 |
| D4 | 6044 | A4* (c/a/b) | 5936/5954/5960 |
| D5 | 6724 | A5 | 5976 |
| | | A6 | 6611 |
| | | A7*(7/7a/7b) | 8335/8341/8369 |
| | | A8 | 9161 |

Table 3 – Description of spliced, partially spliced and unspliced multi-cystronic mRNA, the number of splices per segment and the characteristics of the introns released

| | Length | N.splices | Segment name (mRNA) | INTRONS D1 | A1 | D2 | A2 | D3 | A3 | A4* | A5 | D4 | A6 | D5 | A7* |
|---|---|---|---|---|---|---|---|---|---|---|---|---|---|---|---|
| Stage I: Early transcription | 1.8kb | 1 | nef1 | D1-A7 | | | | | | | | | | | |
| | | 2 | vif1 | D1-A1 | | | | | | | | D4-A7 | | | |
| | | | vpr1 | D1-A2 | | | | | | | | D4-A7 | | | |
| | | | tat1 | D1-A3 | | | | | | | | D4-A7 | | | |
| | | | rev1/2/3 | D1-A4c/a/b | | | | | | | | D4-A7 | | | |
| | | | nef2 | D1-A5 | | | | | | | | D4-A7 | | | |
| | | 3 | vpr2 | D1-A1 | | D2-A2 | | | | | | D4-A7 | | | |
| | | | tat2 | D1-A1 | | D2-A3 | | | | | | D4-A7 | | | |
| | | | rev4/5/6 | D1-A1 | | D2-A4c/a/b | | | | | | D4-A7 | | | |
| | | | nef3 | D1-A1 | | D2-A5 | | | | | | D4-A7 | | | |
| | | | tat3 | D1-A2 | | | | D3-A3 | | | | D4-A7 | | | |
| | | | rev7/8/9 | D1-A2 | | | | D3-A4c/a/b | | | | D4-A7 | | | |
| | | | nef4 | D1-A2 | | | | D3-A5 | | | | D4-A7 | | | |
| | | 4 | tat4 | D1-A1 | | D2-A2 | | D3-A3 | | | | D4-A7 | | | |
| | | | rev10/11/12 | D1-A1 | | D2-A2 | | D3-A4c/a/b | | | | D4-A7 | | | |
| | | | nef5 | D1-A1 | | D2-A2 | | D3-A5 | | | | D4-A7 | | | |
| Stage II: Intermediate | 4.0kb | 1 | vif2 | D1-A1 | | | | | | | | | | | |
| | | | vpr3 | D1-A2 | | | | | | | | | | | |
| | | | tat5 | D1-A3 | | | | | | | | | | | |
| | | | env2/3/4 | D1-A4c/a/b | | | | | | | | | | | |
| | | | env1 | D1-A5 | | | | | | | | | | | |
| | | 2 | vpr4 | D1-A1 | | D2-A2 | | | | | | | | | |
| | | | tat6 | D1-A1 | | D2-A3 | | | | | | | | | |
| | | | env6/7/9 | D1-A1 | | D2-A4c/a/b | | | | | | | | | |
| | | | env5 | D1-A1 | | D2-A5 | | | | | | | | | |
| | | | env10/11/12 | D1-A2 | | | | D3-A4c/a/b | | | | | | | |
| | | | env8 | D1-A2 | | | | D3-A5 | | | | | | | |
| | | 3 | env14/15/16 | D1-A1 | | D2-A2 | | D3-A4c/a/b | | | | | | | |
| | | | env13 | D1-A1 | | D2-A2 | | D3-A5 | | | | | | | |
| Stage III: Late | 9.0kb | 0 | gag/pol | | | | | | | | | | | | |

### 3.3 Translation and protein synthesis model

Ribosome synthesize proteins from messenger RNAs starting scanning from a ribosomal binding site (RBS), as the 5'cap site or an IRES (Internal Ribosome Entry Site). The sequence is scanned from 5' to 3' direction until the first AUG- codon is found. If the first codon is in a weak context, leaky scanning skips this sequence, scanning to the next AUG- codon.

Gag-pol gene contains a programmed ribosomal (-1) frameshift sequence in position [2086], consisting of the sequence -UUUUUA- followed by a -UACAA- cap 9 base-pair long stem-loop, which causes the pause during ribosome translation and shifting position in one nucleotide less. This frameshift produces that a single mRNA sequence can be translated in two different ways: a *gag* p55 precursor [790-2292], encoding capsid proteins, and with a low frequency a *gag-pol* fusion protein [790-5096], encoding capsid and HIV enzymes.

Another -UUUUUA- sequences have been identified in positions [4230] and [6747], corresponding with *pol* and *env* genes. However, it has not been previously reported if these frameshifts have been effective.

There is also a ribosomal SP-stem pause with peptide release in the env gene, before encoding of glycoprotein gp120 (see fig. 6)

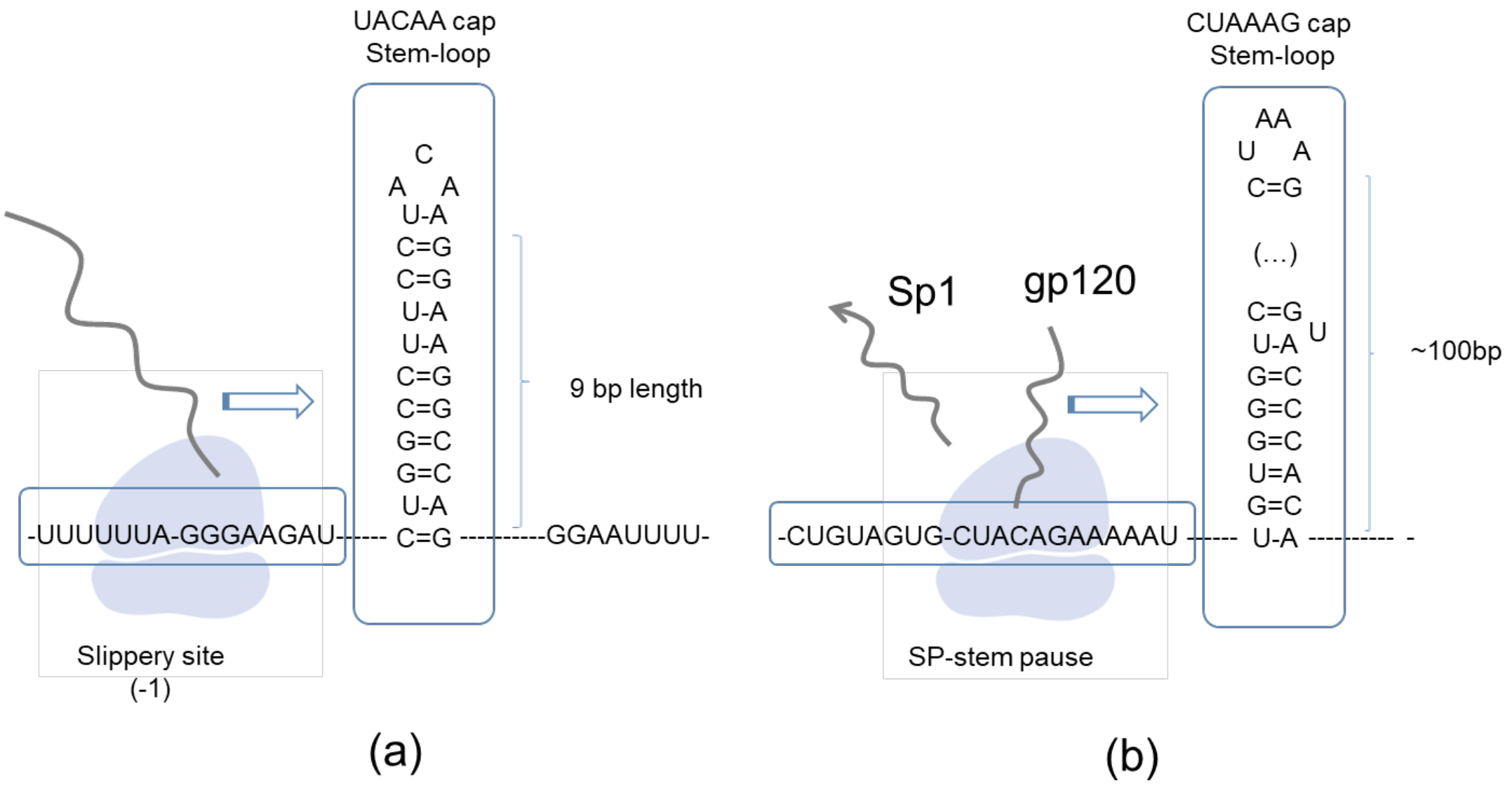

Figure 6 – (a) Ribosomal (-1) frameshift in the gag.pol gene. During translation of gag protein, ribosome pauses at the stem-loop, shifting around 5-10 % times reading-frame in -1 nucleotide and translating gag-pol protein. (b) Ribosomal Sp-pause in the env gene. During translation of env protein, ribosome pauses before the SP-stem loop, releasing the Sp1 signal peptide protein and starting the synthesis of gp120 protein.

### 3.4 Virus maturation

Gag-pol protein precursors emigrate to the cell membrane forcing the budding of the virus in an immature state. After budding, the protease enzyme proteolyzes the gag-pol protein precursors to form the mature state of the virus.

## 4. Results and discussion

### 4.1 Summary

The software HIV.pro simulates the virus replication cycle, modelling specially the transcription process, alternative splicing, ribosomal translation and protein synthesis, and producing several FASTA files with the encoded HIV proteins.

### 4.2 Splicing

The HIV.pro has localised the splicing donor sites (5'ss) with highest intrinsic strength, D1 and D4 (fig. 7), with a slightly difference of +2 nucleotides in the D4 5'ss reference value [10] Further development is still needed to predict how other 5'ss and 3'ss are selected by spliceosome complex proteins, including exonic/intronic enhancers and silencers (EES/ESS/IES/ISS) [49,50] [56].

```
/home/NCBI

########################################################################
 5.5) Splicing:
########################################################################

########################################################################
     Spliceosome model: U1 SnRNP pairs with GU group in intron
     5' splice site sequence  Exon-GU[A/G]AGU...
     Donor splice site
########################################################################

5' splice site: [Exon]-GUGAGU         744
5' splice site: [Exon]-GUAAGU        6046
```

Figure 7 – Splicing simulation with HIV.pro software

## 4.3 Transcription: polyadenylation

The software HIV.pro has identified several polyadenylation cleavage signal Poly(A) in the whole genome, including Poly(A) signals in 5'LTR and 3'LTR [31].

Polyadenylation signals are characterised by the consensus sequence -AAUAAA-, a -CA- sequence and a U/GU rich region. Most of these Poly(A) signals are removed during splicing, within introns. For instance, in the early stage of transcription, intron D1A1 removes poly(A) signals 2, 3, 4 and 5, and intron D4A7* removes poly(A) signals 6, 7 and 8.

These Poly(A) signals are localised after a coding region sequence (CDS), suggesting the hypothesis that poly(A) signals could regulate also some protein expression.

If all these poly(A) signals are significative for starting polyadenylation is still unknown, because at the moment poly(A) signals observed as polyadenylation initiators are those placed in the 5'ltr and 3'ltr region [**Terminator]. However, these unidentified poly(A) signals are followed by several CA sequences, instead of a single sequence CA. In particular, poly(A) 6 has the highest U/GU rich region (see table 4).

Table 4 – Description of the Poly(A) signals for the polyadenylation complex and CDS regions in the HIV-1 (HXB2CG)

| CDS | i | j | Poly(A) signal | | -AAUAAA- sequence position | N. of CA Sequences | GU rich region (%) | U/GU rich Region (%) | Comments |
|---|---|---|---|---|---|---|---|---|---|
| 5'LTR U3 | 1 | 455 | | | | | | | |
| | | | Poly(A) | 1 | 527 | 1 | 26 | 39 | 5'LTR Poly(A) signal |
| 5'LTR U5 | 552 | 633 | | | | | | | Removed with intron D1A1 |
| gag /p17 | 790 | 1185 | | | | | | | |
| | | | Poly(A) | 2 | 1600 | 4 | 6 | 31 | |
| gag /p24 | 1186 | 1881 | | | | | | | |
| gag /p2 | 1882 | 1920 | | | | | | | |
| gag /p7 | 1921 | 2085 | | | | | | | |
| gag /p1 | 2086 | 2133 | | | | | | | |
| | | | Poly(A) | 3 | 2288 | 3 | 10 | 39 | |
| gag /p6 | 2134 | 2292 | | | | | | | |
| pol / PR | 2253 | 2549 | | | | | | | |
| | | | Poly(A) | 4 | 2639 | 2 | 6 | 47 | |
| pol /p51 RT | 2550 | 3869 | | | | | | | |
| | | | Poly(A) | 5 | 4124 | 5 | 8 | 42 | |
| pol /p66 RT | 2550 | 4229 | | | | | | | |
| pol / p15 Rnase H | 3870 | 4229 | | | | | | | |
| | | | Poly(A) | 6 | 4692 | 3 | 10 | 35 | |
| pol /p31 IN | 4230 | 5096 | | | | | | | Removed with intron D4A7* |
| Vif | 5041 | 5619 | | | | | | | |
| Vpr | 5559 | 5850 | | | | | | | |
| Tat exon 1 | 5831 | 6045 | | | | | | | |
| Rev exon 1 | 5970 | 6045 | | | | | | | |
| Vpu | 6062 | 6310 | | | | | | | |
| | | | Poly(A) | 7 | 7290 | 4 | 20 | 54 | |
| | | | Poly(A) | 8 | 7481 | 4 | 12 | 42 | |
| env /gp120 | 6315 | 7757 | | | | | | | |
| Asp | 7373 | 7942 | | | | | | | |
| | | | Poly(A) | 9 | 8070 | 3 | 2 | 39 | |
| env /gp41 | 7758 | 8795 | | | | | | | |
| Tat exon 2 | 8379 | 8469 | | | | | | | |
| Rev exon 2 | 8379 | 8653 | | | | | | | |
| | | | Poly(A) | 10 | 9265 | 4 | 12 | 34 | |
| nef | 8797 | 9417 | | | | | | | |
| | | | Poly(A) | 11 | 9612 | 1 | 14 | 35 | 3'LTR Poly(A) signal |

## 4.4 Protein synthesis

The software HIV-1 has localised all HIV-1 proteins (*gag/pol*, *vpu/env*, *tat*, *rev*, *vif*, *vpr, vpu* and *nef*) and the stage in which they were produced (early, intermediate or late transcription). As HIV mRNAs are multi-cystronic, some proteins are synthesised firstly or in a second place by the ribosome. Table 5 presents the HIV-1 proteins, the number of amino acids and the mRNA sequence that encodes them.

Table 5 – Description of the HIV-1 proteins and the mRNA sequence where

| Transcription stage | Length | Protein | Amino acids | mRNA sequence[a] | |
|---|---|---|---|---|---|
| | | | | 1st protein translated | As secondary protein |
| Stage (I): Early transcription | 0.9kb | nef [b] | 123 | nef1 | - |
| | 1.8kb | nef [b] | 123 | nef2,nef3,nef4,nef5 | Vpr1,vpr2<br>Tat1 to Tat4<br>Rev1 to rev12 |
| | | Vif | 192 | vif1 | - |
| | | Vpr | 78 | vpr1,vpr2 | - |
| | | Tat | Variable[c] | tat1, tat2, tat3, tat4 | Vpr1, vpr2 |
| | | Rev | Variable[d] | rev1 to rev12 | - |
| Stage (II): Intermediate | 4.0kb | Vif | 192 | vif2 | - |
| | | Vpr | 78 | vpr3,vpr4 | - |
| | | tat | 72 | tat5,tat6,tat7,tat8 | - |
| | | rev | 27 | env2,env3,env4<br>env6,env7<br>env9,env10,env11,env12<br>env14,env15,env16 | vpr3,vpr4 |
| | | vpu [b] | 82 | env1<br>env5<br>env8<br>env13 | env2,env3,env4<br>env6,env7<br>env9,env10,env11,env12<br>env14,env15,env16<br>tat5,tat6,tat7,tat8 |
| | | env [b] | 856 | | |
| | | nef [b] | 123 | - | env1,env5,env8,env13 |
| Stage (III): Late transcription | 9.0kb | Gag-pol | 500 | Full length mRNA | - |
| | | | 1435 | Full length mRNA (with -1PRF) | - |

(a) mRNA sequences defined as Purcell and Martin (1993) [7]

(b) Ribosomal leaky scanning

(c) If mRNA is created with exon 7, protein with 224aa; if exon 7a, with 222aa and with exon 7b, 89. If exon 6 is included, *tat* has 76aa.

(d) If mRNA is created with exon 7, protein with 25aa; if exon 7a, with 52aa and with exon 7b, 119. If exon 6 is included, *rev* has28aa.

### 4.4.1 *Nef*

Nef1 sequence is the shortest mRNA produced during all the process, encoding only *ne*f protein. *Nef* protein is translated in the early stage of transcription with exon 7, after a ribosomal leaky scanning. *Nef* length is constant during all the process, not influenced by the A7* 3'ss.

### 4.4.2 *Tat/rev*

*Tat* and *rev* proteins are encoded in the virus genome between spliced exons. The AUG- starting codon of tat protein is in exon 4 and the end codon in exon 7, whereas *rev* protein has starting codon in exon 4cab and end codon in exon 7ab. As exons 4cab and 7 have variable 3'ss acceptor sites, the released intron and coding sequence have different lengths, producing several fusion proteins.

In the case of a completely spliced 1.8kb mRNA, exon 7 produces the longest *tat* protein, with 224aa and the smallest *rev*, with 25aa, whereas exon 7b produces the opposite, *tat* with 89aa and *rev* with 119aa.

As far as splicing is inhibited, in the case of 4.0kb mRNA, *tat* and *rev* reach constant lengths of 72aa and 27aa.

Exon 7 is determinant in the length of both proteins, *tat* and *rev*, whereas exon 4cab is not determinant in the length of protein *rev*. Exon 6 produces tat/rev proteins with constant length, independent of A7 value.

If exon 6 is included (A6-D5), tat protein has a length of 76aa and rev of 28aa. These expressions of *tat*/*rev* proteins have been previously described in some studies as *tev* protein or fusion proteins [52-53].

As these proteins are absent or in a low frequency in most of the HIV analysis, splicing happens inefficiently in early stages, until the synthesis of tat/rev proteins are stable.

Annexes B and C describes the tat and rev splicing and the FASTA files of the proteins.

### 4.4.3 *Vpu*

*Vpu* is translated as a secondary protein with *tat5* to *tat8* mRNA segments. Although *vpu* and *env* genes share a common nucleotide region, both proteins are independently translated in different mRNA.

### 4.4.4 *Env*

A ribosome shunting regulates translation of *env* protein, produced by the sequence -AUG-UAA-. After this shunting, ribosome continues scanning *env* gene until founding an Sp- stem-loop, placed at the beginning of gp120 glycoprotein, making a pause long enough to disrupt the polypeptide synthesis, releasing signal peptide 1 (Sp1) protein and continuing translating gp120 and gp41 glycoprotein (gp160) [6]. After translation, glycoprotein gp160 binds to the cell membrane. Other regions are encoded in env gene, as the antisense protein (asp), although there is still no evidence of being translated during the HIV virus cycle. Figure 8 and table 6 presents a scheme of the characteristic regions of *env* gene.

Several studies have reported the placement of the binding site of the gp120 to CD4 cells in some region between the conformation loops, although it is still unclear.

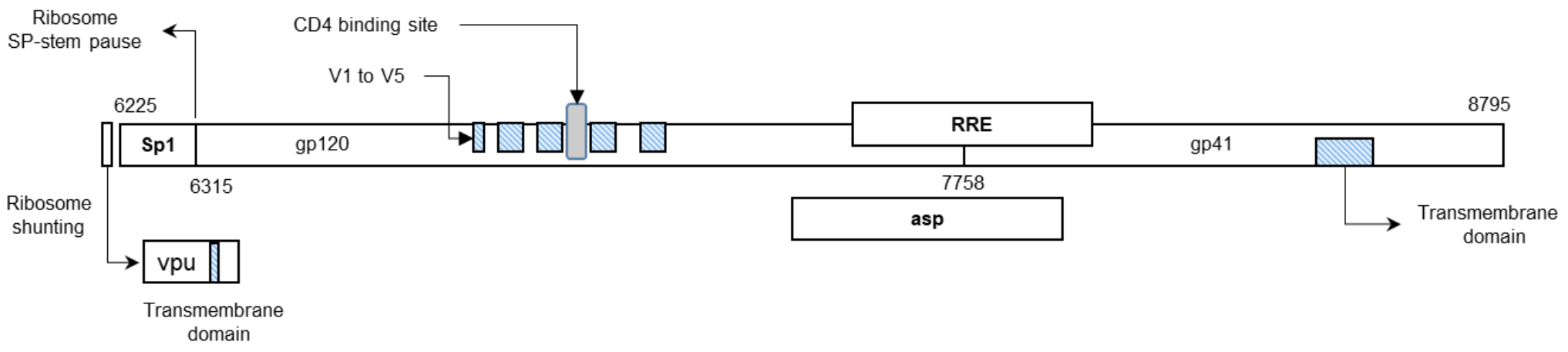

Figure 8 – Scheme of env gene, starting with a ribosome shunting, SP-stem pause and translation of gp120/gp41 proteins

Table 6 - Description of *env* proteins [31]

| | | i | j | Description |
|---|---|---|---|---|
| Sp1 | - | 6225 | 6314 | Signal peptide 1 before stem loop |
| gp120 [6315-7757] | V1 | 6615 | 6692 | V1 loop |
| | V2 | 6696 | 6812 | V2 loop |
| | V3 | 7110 | 7217 | V3 loop |
| | V4 | 7377 | 7478 | V4 loop |
| | VBS-CD4 | 7479 | 7601 | Binding site CD4 [(a)] |
| | V5 | 7602 | 7634 | V5 loop |
| gp120-gp41 | RRE | 7710 | 8061 | Rev responsive element (RRE), binding site of *rev* protein |
| | Asp | 7942 | 8795 | Antisense protein |
| gp41 [7758-8795] | Gp41 TMD | 8277 | 8336 | Transmembrane domain of gp41 glycoprotein |

[(a)] Other studies suggest HIV binding site to CD4 receptor between V2/V3 [64]

## 4.4.5 *Gag-pol* proteins

At the end of the transcription process, spliceosome activity is inhibited and full-length 9.0kb mRNA are exported to the ribosome for translation, encoding gag-pol protein. As commented, gag-pol can be translated into two different polypeptide chains with a different probability, due to the ribosomal frameshift (table 7). These polypeptides chains reorganise and bind to the cell membrane to conform an immature provirus which buds out of the cell. HIV virus maturation is still unknown, being protease enzyme decisive in this process.

Table 7 - Description of gag-pol protein: precursor p55 and gag-pol fusion protein p160 [47]

| P(%) | Polypeptide | RF | i | j | Aa | Proteins | Function | Description |
|---|---|---|---|---|---|---|---|---|
| 90-95 | Gag precursor (p55) | +1 | 790 | 1185 | 132 | p17 (MA) | Structural | Spherical matrix to support lipid membrane |
| | | | 1186 | 1881 | 232 | P24 (CA) | | Capsid |
| | | | 1882 | 1920 | 13 | p2 | Connector | Spacer peptide 1 |
| | | | 1921 | 2085 | 55 | p7 (NC) | Structural | Nucleocapsid |
| | | | 2086 | 2133 | 16 | p1 | Connector | Spacer peptide 2 |
| | | | 2134 | 2292 | 53 | p6 | Assembly | Facilitates ESCRT-dependent budding and allows incorporation of *vpr* into virus |
| 5-10 | Gag-pol fusion (p160) | +1 | 790 | 2085 | 432 | P17/p24/p2/p7 | Structural | Gag proteins |
| | | +3 | 2085 | 2252 | 56 | TFP | | Trans-frame fusion protein |
| | | | 2253 | 2549 | 99 | PR/p10 | Enzymes | Protease |
| | | | 2550 | 3869 | 440 | RT/p51 | | Reverse transcriptase |
| | | | 2550 | 4229 | 560 | RT/p66 | | Reverse transcriptase |
| | | | 3870 | 4229 | 120 | Rnase H / p15 | | Ribonuclease H |
| | | | 4230 | 5096 | 289 | IN/p31 | | Integrase |

## 4.4.6 Other ORF encoding proteins

A total number of 171 open reading frames (ORF) has been recognised in the whole genome, including the HIV-1 proteins previously described.

Some of these ORF encode proteins with a small number of amino acids that undergo real translation, that is, short ORF-encoded polypeptides (sEPs), being unclear if they are significative during the virus replication process. Other ORF remain without translation because the starting codon is in a reading frameshift not scanned by the ribosome. There are still some proteins that could be translated and remain unidentified, because they are degraded during the replication process or because as being the second or third ORF in a mRNA translation are produced in a low frequency. Table 8 summarizes all these ORF and the protein that should be expressed.

Table 8 - Description of ORF with unidentified proteins in the 1.8kb, 4.0kb and 9.0kb mRNA

| | mRNA | Start codon | End codon | Aa | Protein | Comments |
|---|---|---|---|---|---|---|
| 1.8kb | Nef1,nef2 Rev1 to rev12 Vpr1,vpr2 | 9313 | 9418 | 34 | MDDPEREVLEWRFDSRLAFHHVARELHPEYFKNCX | After *nef* protein |
| | Tat 1 to tat4 | 8832 | 8868 | 11 | MAYCKGKNETSX | |
| | | 9216 | 9243 | 8 | MVLQASTSX | Also in [131-158] |
| 4.0kb | Vif2 | 5641 | 5686 | 14 | MKLLDIFLGFGSMAX | |
| | | 5698 | 5740 | 13 | MKLMGILGQEWKPX | |
| | | 6343 | 6478 | 44 | MGYLCGRKQPPLYFVHQMLKHMIQRYIMFGPHMPVYPQTPTHKKX | |
| | | 6925 | 6988 | 20 | MEQDHVQMSAQYNVHMELGQX | |
| | | 8056 | 8149 | 30 | MLVGVINLWNRFGITRPGWSGTEKLTITQAX | |
| | | 9313 | 9418 | 34 | MDDPEREVLEWRFDSRLAFHHVARELHPEYFKNCX | As found in 1.8kb mRNA |
| | Vpr3,vpr4 Env1 to env16 | 8832 | 8868 | 11 | MAYCKGKNETSX | As found in 1.8kb mRNA |
| | | 9216 | 9243 | 8 | MVLQASTSX | As found in 1.8kb mRNA |
| | Tat5 to tat8 | 6557 | 6608 | 16 | MGSKPKAMCKINPTLCX | |
| | | 6938 | 6980 | 13 | MYKCQHSTMYTWNX | |
| | | 7475 | 7556 | 26 | MQNKTNYKHVAESRKSNVCPSHQWTNX | |
| | | 7556 | 7604 | 15 | MFIKYYRAAINKRWWX | |
| | | 8915 | 8972 | 18 | MEQSQVAIQQLPMLLVPGX | |
| | | 9308 | 9341 | 10 | MGWMTRREKCX | |
| 9.0kb | Gag/pol | 1251 | 1409 | 52 | MGKSSRREFQPRSDTHVFSIIRRSHPTRFKHHAKHSGGTSSSHANVKRDHQX | Reading frame +3 |

## 5. Conclusions

The software HIV.pro identifies with high accurateness the location and the amino acid composition of the proteins in the HIV1(HXB2) genome, producing the FASTA files for further analysis. All HIV-1 proteins have been localised in the genome sequence, making special attention to the tat/rev proteins. In this case, it has been observed that the length of these proteins depends on the splicing process, especially on the 3'ss acceptor site A7, where the U2 snRNP spliceosome protein binds during splicing. Furthermore, other ORF that could be translated into proteins have been identified.

The software has identified other features, as ribosomal frameshifting sequences and cleavage signals for polyadenylation, needing special focus and analysis.

Although this type of software has been extensively developed, with different kind of features and analyses, as for identifying secondary structures or enhancers/silencers sequences (ESE/ESS/ISS), it can be applied to any type of genome, allowing further implementations, such as splicing prediction or mutations. Other HIV-1 genome sequences and subtypes will be analysed in future studies.

The detailed analysis of the HIV virus replication and the characterisation of virus proteomics are important for identifying which antigens are presented by macrophages to CD4 cells, for localizing reactive epitopes or for creating transfer vectors to develop new HIV vaccines and effective therapies, where intensive research is still needed.

## 6. References

[1] Gallo RC, Salahuddin SZ, Popovic M, Shearer GM, Kaplan M, Hayes BF, Parker TJ, Redfneld R, Oleske J, Safai B, White G, Foster P and Markham PD (1984) Science (1984) 224, 500-503

[2] Levy JA, Hoffman AD, Kramer SM, Lanois JA, Shimabukuro JM and Oskiro LS (1984) Science 225, 840-842

[3] Barre-Sinoussi F, Chermann JC, Rey F, Nugeybe MT, Chamaret S, Gruest J, Dauguet C, Axlar-Blin C, Vezinet-Brun F, Rouzioux C, Rozenbaum W and Montagnier L (1983) Science 220, 868-870

[4] Dendritic cells in progression and pathology of HIV infection. O Manches, D Frleta and N Bhardwaj. Trends Immunol. 2014 Mar; 35(3): 114–122. doi: 10.1016/j.it.2013.10.003

[5] HIV-1: Molecular Biology and Pathogenesis Viral Mechanisms, Second Edition (2007). Kuan-Teh Jeang Volume 55, Pages 1-467

[6] Architecture and Secondary Structure of an Entire HIV-1 RNA Genome. Watts et al. Nature (2009) 460(7256): 711-716. doi:10.1038/nature08237

[7] Alternative Splicing of Human Immunodeficiency Virus Type 1 mRNA Modulates Viral Protein Expression, Replication, and Infectivity. JOURNAL OF VIROLOGY, Nov. 1993. American Society for Microbiology. D Purcell and M Martin - P.6365-63780022-538X/93/116365-14$02.00/0 –

[8] Alternative splicing: regulation of HIV-1 multiplication as a target for therapeutic action. J Tazi, N Bakkour, V Marchand, L Ayadi, A Aboufirassi, C Branlant. FEBS Journal (2010) v277(4)

[9] Dynamic regulation of HIV-1 mRNA populations analysed by single-molecule enrichment and long-read sequencing. KE Ocwieja et al. Nucleic Acids Res. 2012 Nov; 40(20): 10345–10355. doi: 10.1093/nar/gks753

[10] Splicing of human immunodeficiency virus RNA is position-dependent suggesting sequential removal of introns from the 5′ end. JBH Wodrich and H Kräusslich. Nucleic Acids Res. 2005; 33(3): 825–837. doi: 10.1093/nar/gki185

[11] Cloning and functional analysis of multiply spliced mRNA species of human immunodefiency virus type 1. S Schwartz, BK Felber, DM Benko, EM Fenyö and GN Pavlakis. J Virol. 1990 Jun; 64(6): 2519-2529

[12] Transcriptional and Posttranscriptional Regulation of HIV-1 Gene Expression. J Karn and C Martin Stoltzfus. Cold Spring Harb Perspect Med. 2012 Feb; 2(2): a006916. doi: 10.1101/cshperspect.a006916

[13] Characterizing HIV-1 Splicing by Using Next-Generation Sequencing. A Emery, S Zhou, E Pollom and R Swanstrom. J Virol. 2017 Mar 15; 91(6): e02515-16. doi: 10.1128/JVI.02515-16

[14] K Mayo, D Huseby, J McDermott, B Arvidson, L Finlay, E Barklis. Retrovirus capsid protein assembly arrangements, J. Mol. Biol., 325(2003), pp. 225–237

[15] An atomic model of HIV-1 capsid-SP1 reveals structures regulating assembly and maturation. FKM Schur, M Obr, WJH Hagen, W Wan, AJ Jakobi, JM Kirkpatrick, C Sachse, HG Kräusslich, JAG Briggs. Science 29 Jul 2016: Vol. 353, Issue 6298, pp. 506-508 DOI: 10.1126/science.aaf9620

[16] Three-dimensional structure of the human immunodeficiency virus type 1 matrix protein. Massiah MA, Starich MR, Paschall C, Summers MF, Christensen AM, Sundquist WI (1994) J.Mol.Biol. 244: 198-223

[17] Organization of Immature Human Immunodeficiency Virus Type 1. T Wilk, I Gross, BE Gowen, T Rutten, F de Haas, R Welker, HG Kräusslich, P Boulanger, SD Fuller (2001) Journal of Virology 75 (2) p. 759–771. doi: 10.1128/JVI.75.2.759–771.2001

[18] Yang H, Jossinet F, Leontis N, Chen L, Westbrook J, Berman HM, Westhof E. Tools for the automatic identification and classification of RNA base pairs. *Nucleic Acids Research* (2003) 31.13: 3450-3460.

[19] Structure of an HIV gp120 envelope glycoprotein in complex with the CD4 receptor and a neutralizing human antibody. Kwong et al. Nature (1998) 393(6686): 648-659

[20] P Prabakaran, AS Dimitrov, TR Fouts, DS Dimitr. Structure and Function of the HIV Envelope Glycoprotein as entry mediator, vaccine immunogen and target for inhibitors. Advances in Pharmacology (2007) 55, 33-97

[21] Poliovirus, Pathogenesis of Poliomyelitis and Apoptosis. Blondel B, Colbere-Garapin F, Couderc T, Wirotius A and Guivel-Benhassime F. CTMI (2005) 289:25-56 doi: 10.1007/3-540-27320-4_2

[22] Numbering Positions in HIV Relative to HXB2CG, in the database compendium. Korber et al. Human Retroviruses and AIDS (1998)

[23] HIV-1 Replication and the Cellular Eukaryotic Translation Apparatus. S Guerrero, J Batisse, C Libre, S Bernacchi, R Marquet and JC Paillart. Viruses 2015, 7, 199-218; doi:10.3390/v7010199

[24] Env and Vpu proteins of human immunodeficiency virus type 1 are produced from multiple bicistronic mRNAs. Schwartz S, Felber BK, Fenyö EM, Pavlakis GN. J Virol. 1990 Nov; 64(11):5448-56.

[25] Transcriptional and Post-transcriptional Regulation of HIV-1 Gene Expression. J Karn and C Martin Stoltzfus. Cold Spring Harb Perspect Med. 2012 Feb; 2(2): a006916. doi: 10.1101/cshperspect.a006916

[26] M Hill, G Tachedjian and J Mak. The Packaging and Maturation of the HIV-1 Pol Proteins. Current HIV Research 3(1):73-85. February 2005 doi: 10.2174/1570162052772942

[27] AML Lever. HIV-1 RNA Packaging. Advances in Pharmacology. Volume 55, 2007, Pages 1-32

[28] IRESite: the database of experimentally verified IRES structures (www.iresite.org) M Mokrejš et al. Nucleic Acids Research, Volume 34, Issue suppl_1, 1 January 2006, Pages D125–D130, https://doi.org/10.1093/nar/gkj081

[29] Searching for IRES. Stephen D. Baird, Marcel Turcotte, Robert G. Korneluk and Martin Holcik. RNA. 2006 Oct; 12(10): 1755–1785. doi: 10.1261/rna.157806

[30] IRESdb: The Internal Ribosome Entry Site database. Nucleic Acids Res. 2003 Jan 1;31(1):427-8. Bonnal, Boutonnet C, Prado-Lourenço L, Vagner S.

[31] HIV Sequence Compendium 2018. Brian Foley, Thomas Leitner, Cristian Apetrei, Beatrice Hahn, Ilene Mizrachi, James Mullins, Andrew Rambaut, Steven Wolinsky, and Bette Korber editors. 2018. Publisher: Los Alamos National Laboratory, Theoretical Biology and Biophysics, Los Alamos, New Mexico. LA-UR-18-25673.

[32] HIV: Cell Binding and Entry. CB Wilen, JC Tilton and RW Doms. Cold Spring Harb Perspect Med. 2012 Aug; 2(8) doi: 10.1101/cshperspect.a006866

[33] HIV-1 Ribonuclease H: Structure, Catalytic Mechanism and Inhibitors. GL Beilhartz and M Götte. Viruses (2010) 2(4) 900-926. doi: 10.3390/v2040900

[34] HIV-1 Reverse Transcription. W Hu and SH Hughes. Cold Spring Harb Perspect Med (2012) 2(10) doi: 10.1101/cshperspect.a006882

[35] HIV DNA Integration. R Craigie and FD Bushman. Cold Spring Harb Perspect Med. 2012 Jul; 2(7): a006890. doi: 10.1101/cshperspect.a006890

[36] RNA Polymerase-Promoter Interactions: the Comings and Goings of RNA Polymerase. PL deHaseth, ML Zupancic and MT Record. Jr. J Bacteriol. 1998 Jun; 180(12): 3019–3025.

[37] Jones KA, Luciw PA & Duchange N (1988). Structural arrangements of transcription control domains within the 5'-untranslated leader regions of the HIV-1 and HIV-2 promoters. Genes Develop 2, 1101-1114.

[38] Garcia JA, Harrich D, Soultanakis E, Wu F, Mitsuyasu R and Gaynor RB (1989). Human immunodeficiency virus type 1 LTR TATA and TAR region sequences required for transcriptional regulation. EMBO J. 8, 765-778.

[39] Olsen HS and Rosen CA. Contribution of the TATA motif to Tat-mediated transcriptional activation of the human immunodeficiency virus gene expression (1992). J. Virol.66, 5594-5597.

[40] The strength of the HIV-1 3' splice sites affects Rev function. S Kammler, M Otte, I Hauber, J Kjems, J Hauber and H Schaal. Retrovirology. 2006; 3: 89. doi: 10.1186/1742-4690-3-89

[41] The HIV-1 Tat Protein: Mechanism of Action and Target for HIV-1 Cure Strategies. Rice AP. Curr Pharm Des (2017) 23 (28): 4098-4102. doi: 10.2174/1381612823666170704130635.

[42] Mechanism of translation of monocistronic and multicistronic human immunodeficiency virus type 1 mRNAs. S Schwartz, BK Felber, and GN Pavlakis. Mol Cell Biol. 1992 Jan; 12(1): 207–219

[43] The Human Immunodeficiency Virus Type 1 Ribosomal Frameshifting Site Is an Invariant Sequence Determinant and an Important Target for Antiviral Therapy. P Biswas, X Jiang, AL Pacchia, JP Dougherty and SW Peltz. J Virol. 2004 Feb; 78(4): 2082–2087. doi: 10.1128/JVI.78.4.2082-2087.2004

[44] Characterization of ribosomal frameshifting in HIV-1 gag-pol expression. Jacks T, Power MD, Masiarz FR, Luciw PA, Barr PJ, Varmus HE. Nature. 1988 Jan 21;331(6153):280-3.

[45] Retroviral proteases. BM Dunn, MM Goodenow, A Gustchina and A Wlodawer. Genome Biol. 2002 3(4)

[46] Yu-Dong Cai and Kuo-Chen Chou. Artificial neural network model for predicting HIV protease cleavage sites in protein. Advances in Engineering Software (1998) 29 (2) 119-128.

[47] Bioafrica. Bioinformatics & Genomics for Health and Life Sciences in Africa. (http://mail.bioafrica.net/proteomics)

[48] The biased nucleotide composition of the HIV genome: a constant factor in a highly variable virus. AC van der Kuyl and B Berkhout. Retrovirology. 2012; 9: 92. doi: 10.1186/1742-4690-9-92

[49] Insights into the selective activation of alternatively used splice acceptors by the human immunodeficiency virus type-1 bidirectional splicing enhancer. C Asang, I Hauber, H Schaal. (2008). Nucleic Acids Res, 1-14

[50] Balanced splicing at the Tat-specific HIV-1 3'ss A3 is critical for HIV-1 replication. S Erkelenz, F Hillebrand, M Widera, S Theiss, A Fayyaz, D Degrandi, K Pfeffer and H Schaal. Retrovirology. 2015; 12: 29. doi: 10.1186/s12977-015-0154-8

[51] Mechanisms of HIV Transcriptional Regulation and Their Contribution to Latency. GM Schiralli Lester and AJ Henderson. Mol Biol Int. 2012. doi: 10.1155/2012/614120

[52] A novel human immunodeficiency virus type 1 protein, tev, shares sequences with tat, env, and rev proteins. Benko DM, Schwartz S, Pavlakis GN, Felber BK. J Virol. 1990 Jun;64(6):2505-18

[53] Unusual Fusion Proteins of HIV-1. S Langer and D Sauter. Frontiers in Microbiology 2017 vol 7 art 2152. doi: 10.3389/fmicb.2016.02152

[54] MF Hagan, Modelling viral capsid assembly, Adv. Chem. Phys., 155(2014), pp. 1–68

[55] OM Elrad, MF Hagan. Encapsulation of a polymer by an icosahedral virus. Phys.Biol., 7(2010), pp. 045003:1–17

[56] C Asang, I Hauber and H Schaal. Insights into the selective activation of alternatively used splice acceptors by the human immunodeficiency virus type-1 bidirectional splicing enhancer. Nucleic Acids Res (2008) 1-14

[57] M Balasubramaniam, EO Freed. New insights into HIV assembly and trafficking, Physiology, 26(2011), pp. 236-251

[58] Complete nucleotide sequence of the AIDS virus, HTLV-III. Nature 313 (6000), 277-284 (1985)

[59] Wain-Hobson S, Vartanian JP, Henry M, Chenciner N, Cheynier R, Delassus S, Martins LP, Sala M, Nugeyre MT, Guetard D. LAV revisited: origins of the early HIV-1 isolates from Institut Pasteur. Science 252(5008); 961-5 (1991)

[60] HIV Sequence Databases. C Kuiken, B Korber and RW Shafer. AIDS Rev. 2003; 5(1): 52–61.

[61] A Composite Polyadenylation Signal with TATA Box Function. N Paran, A Ori, I Haviv, and Y Shaul. Mol Cell Biol. 2000 Feb; 20(3): 834–841.

[62] The activity of the HIV-1 IRES is stimulated by oxidative stress and controlled by a negative regulatory element. K Gendron, G Ferbeyre, N Heveker and L Brakier-Gingras. Nucleic Acids Res. 2011 Feb; 39(3): 902–912. doi: 10.1093/nar/gkq885

[63] Regulation of alternative RNA splicing by exon definition and exon sequences in viral and mammalian gene expression. Z Zheng. J Biomed Sci. J Biomed Sci. 2004; 11(3): 278–294. doi: 10.1159/000077096

[64] Adhesion and fusion efficiencies of human immunodeficiency virus type 1 (HIV-1) surface proteins. TM Dobrowsky, SA Rabi, R Nedellec, BR Daniels, JI Mullins, DE Mosier, RF Siliciano and D Wirtz. Scientific Reports 3, 3014 (2013)

[65] Reviving an old HIV-1 gene: The HIV-1 antisense protein. Torresilla C, Mesnard JM, Barbeau B. Current HIV Res. (2015); 13(2): 117-24

[66] Concomitant emergence of the antisense protein gene of HIV-1 and of the pandemic. Cassan E, Arigon-Chifolleau AM, Mesnard JM, Gross A, Gascuel O. Proc Natl Acad Sci USA (2016) Oct 11; 113(41): 11537-11542

Annex A. HIV-1 genome sequence: type HXB2CG

The NCBI (National Center for Biotechnology Information) has published the last version of the HIV-1 complete genome (Aug-2018). This sequence is based on a compendium of isolated and/or fractional genomes, aligned to an initial alignment sequence [31] [50]

The NCBI REFSEQ is a reference genome sequence for the retrovirus HIV-1 subtype (HXB2CG), last version NC_001802.1 from Aug-2018. The subtype HXB2 is a clone from the French isolate LAI (formerly LAV-BRU) [59]. The sequence includes a total of 9181 nucleotides, arranged mainly in structural, regulatory and auxiliary genes. The sequence has been modified including the regions 5'LTR u5 (454bp) and 3'LTR u5 (84bp), from version K03455.1 (2002), for a total length of 9719 nucleotides.

Table A1 and figure A1 shows a detailed description of the HIV-1 virus 5’ltr and 3’ltr.

Table A1 – Regulatory elements in the untranslated region 5’ltr and 3’ltr

| | | 5’LTR | 3’LTR | Nucleotide sequence |
|---|---|---|---|---|
| LTR U3 | | [1:455] | [9086:9540] | - |
| TCF-1 alpha | | [315:329] | [9400:9414] | uacuucaagaacugc |
| NF | NF-k-B-II | [350:359] | [9435:9444] | gggacuuucc |
| | NF-k-B-I | [364:373] | [9449:9458] | gggacuuucc |
| Sp1 | Sp1-III | [375:386] | [9462:9471] | gggaggcguggc |
| | Sp1-II | [388:397] | [9473:9482] | ugggcgggac |
| | Sp1-I | [398:408] | [9483:9493] | uggggaguggc |
| TATA box | | [427:431] | [9512:9516] | uauaa |
| TAR element | | [453:513] | [9538:9599] | ugggucucucugguuagaccagaucugagccugggagcucucuggcuaacuagggaaccca |
| Poly(A) signal | | [527:532] | [9612:9617] | aauaaa |

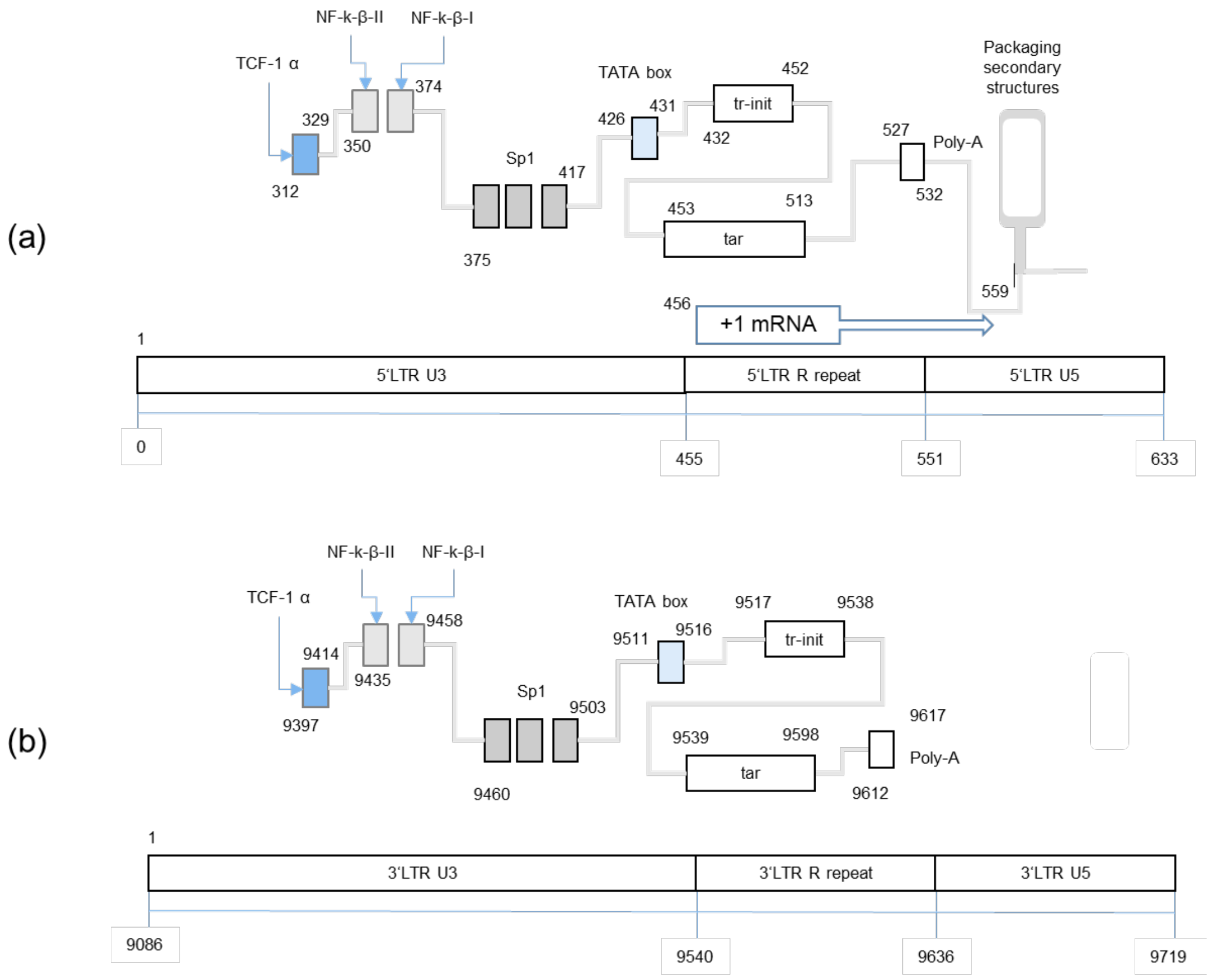

Figure A1 – (a) Description of the main elements of the 5’ LTR promoter region of the RNAPII and (b) 3’ LTR of the HIV-1 genome. As observed, both regions include repetitive elements, as the transcription factor NF, Sp1 and TATA box [61].

## Annex B. Splicing of tat and rev HIV-1 proteins

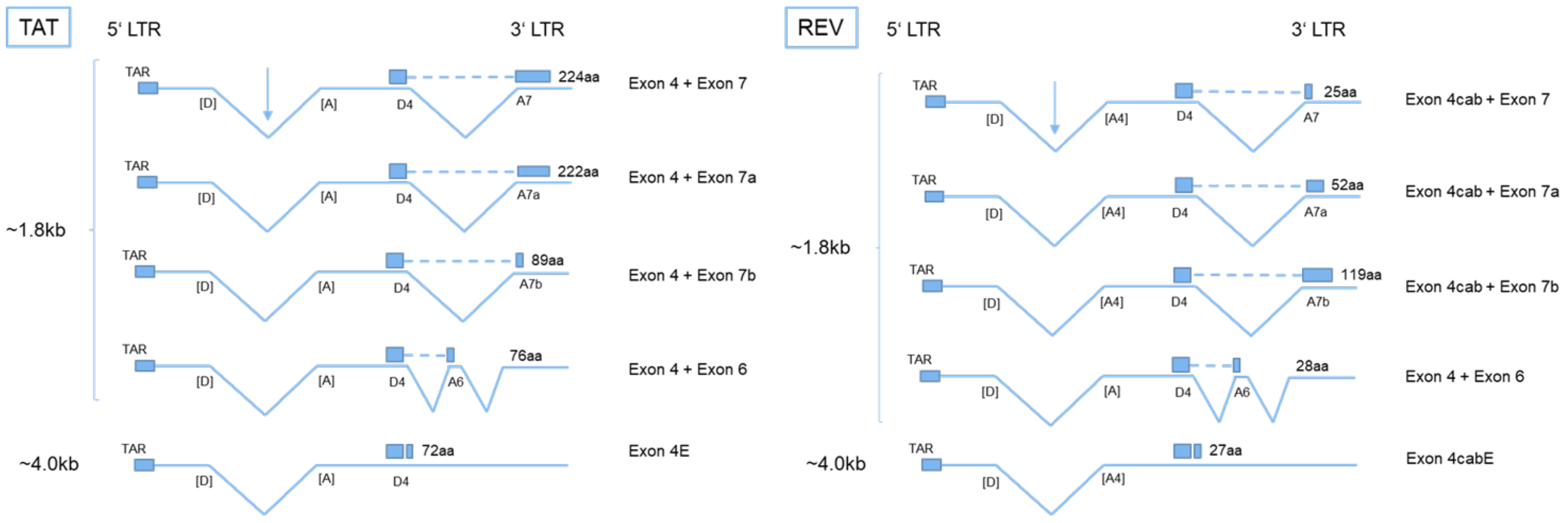

Figure B1 – Splicing of the *tat* and *rev* genes [52-53]: (a) *tat* protein is formed with a fragment of exon 4 and another from exon 7, 7a or 7b, which has a variable length, depending on the 3' splicing acceptor site A7. Furthermore, in the case of exon 7b, there is a change in the reading frame between exons, reducing considerably *tat* length. Another case, if exon 6 is present, *tat* length is constant (b) *Rev* protein is formed with a fragment of exon 4c, 4a or 4b and another from exon 7, 7a or 7b. Although 3'ss A4 is variable, is not determinant for rev length. If exon 6 is present, *rev* length is constant

## Annex C. FASTA files

```
TAT protein
1.8kb

 *********************************************
 tat1 /.../tat4
 *********************************************
 >ncbi|            1 |            1 |          344         1016          675 |          224
 MEPVDPRLEPWKHPGSQPKTACTNCYCKKCCFHCQVCFITKALGISYGRKKRRQRRRAHQNSQTHQASLSKLVNRVRQGYSPLSFQTHLP
 TPRGPDRPEGIEEEGGERDRDRSIRLVNGSLALIWDDLRSLCLFSYHRLRDLLLIVTRIVELLGRRGWEALKYWWNLLQYWSQELKNSAV
 SLLNATAIAVAEGTDRVIEVVQGACRAIRHIPRRIRQGLERILLX

 *********************************************
 tat1.7a/.../tat4.7a
 *********************************************
 >ncbi|            1 |            1 |          344         1010          669 |          222
 MEPVDPRLEPWKHPGSQPKTACTNCYCKKCCFHCQVCFITKALGISYGRKKRRQRRRAHQNSQTHQASLSKHRVRQGYSPLSFQTHLPTP
 RGPDRPEGIEEEGGERDRDRSIRLVNGSLALIWDDLRSLCLFSYHRLRDLLLIVTRIVELLGRRGWEALKYWWNLLQYWSQELKNSAVSL
 LNATAIAVAEGTDRVIEVVQGACRAIRHIPRRIRQGLERILLX

 *********************************************
 tat1.7b/.../tat4.7b
 *********************************************
 >ncbi|            1 |            1 |          344          611          270 |           89
 MEPVDPRLEPWKHPGSQPKTACTNCYCKKCCFHCQVCFITKALGISYGRKKRRQRRRAHQNSQTHQASLSKHRFRPTSQPRGDPTGPKEX

 *********************************************
 tat1.6/.../tat4.6
 *********************************************
 >ncbi|            1 |            1 |          468          698          230 |           76
 MEPVDPRLEPWKHPGSQPKTACTNCYCKKCCFHCQVCFITKALGISYGRKKRRQRRRAHQNSQTHQASLSKQSALIX


4.0kb

*********************************************
 tat5/6/7
 *********************************************
 >ncbi|            1 |            1 |          344          560          219 |           72
 MEPVDPRLEPWKHPGSQPKTACTNCYCKKCCFHCQVCFITKALGISYGRKKRRQRRRAHQNSQTHQASLSKQX
```

Figure C1 – FASTA file for *tat* proteins: 1.8kb and 4.0kb mRNAs (not expressed at 9.0kb)

```
Rev protein
1.8kb

*********************************************
rev11/.../10
*********************************************
>ncbi|          1 |           1 |         300         375          78 |           25
MAGRSGDSDEELIRTVRLIKLLYQSX

*********************************************
rev1.7a/.../10.7a
*********************************************
>ncbi|          1 |           1 |         300         456         159 |           52
MAGRSGDSDEELIRTVRLIKLLYQSIELGRDIHHYRFRPTSQPRGDPTGPKEX

*********************************************
rev1.7b/.../10.7b
*********************************************
>ncbi|          1 |           1 |         300         657         360 |          119
MAGRSGDSDEELIRTVRLIKLLYQSIVSDPPPNPEGTRQARRNRRRRWRERQRQIHSISERILGTYLGRSAEPVPLQLPPLERLTLDCNE
DCGTSGTQGVGSPQILVESPTVLESGTKEX

*********************************************
rev1.6/.../10.6
*********************************************
>ncbi|          1 |           1 |         424         510          86 |           28
MAGRSGDSDEELIRTVRLIKLLYQSKVHX

4.0kb

*********************************************
env16/15/14-12/11/10/9-7/6-4/3/2 + vpr3/vpr4
*********************************************
>ncbi|          1 |           1 |         448         529          84 |           27
MAGRSGDSDEELIRTVRLIKLLYQSSKX
```

Figure C2 – FASTA files for *rev* proteins: 1.8kb and 4.0kb mRNAs (not expressed at 9.0kb)